%% file: main.tex
\documentclass{article}
\usepackage{spconf,amsmath,graphicx}
\usepackage{amssymb}

\usepackage{tikz}
\usetikzlibrary{shapes, calc, fit}
\usepackage{fix-cm}

\usepackage{hyperref}

\let\OLDthebibliography\thebibliography
\renewcommand\thebibliography[1]{
  \OLDthebibliography{#1}
  \setlength{\parskip}{1pt plus 0.3ex minus 0.3ex}
  \setlength{\itemsep}{1pt plus 0.3ex minus 0.3ex}
}

\usepackage{anyfontsize}
\usepackage{subfig}

\def\L{{\cal L}}
\def\E{{\mathbb{E}}}
\newcommand{\norm}[1]{\left\lVert#1\right\rVert}

\usepackage{bm}

\title{Waveform generation for text-to-speech synthesis using pitch-synchronous multi-scale generative adversarial networks}
\name{
\begin{tabular}{c}
Lauri Juvela$^1$, Bajibabu Bollepalli$^1$,  Junichi Yamagishi$^{2,3}$, Paavo Alku$^1$ \\ 
\end{tabular}
}
\address{
        $^1$Aalto University, Finland \hspace{10mm}
        $^2$National Institute of Informatics, Japan \\ 
        $^3$The Centre for Speech Technology Research, University of Edinburgh, United Kingdom \\
}
\begin{document}
\ninept
\maketitle
\begin{abstract}
The state-of-the-art in text-to-speech synthesis has recently improved considerably due to novel neural waveform generation methods, such as WaveNet. However, these methods suffer from their slow sequential inference process, while their parallel versions are difficult to train and even more  expensive computationally.
Meanwhile, generative adversarial networks (GANs) have achieved impressive results in image generation and are making their way into audio applications; parallel inference is among their lucrative properties. By adopting recent advances in GAN training techniques, this investigation studies waveform generation for TTS in two domains (speech signal and glottal excitation).
%
Listening test results show that while direct waveform generation with GAN is still far behind WaveNet, a GAN-based glottal excitation model can achieve quality and voice similarity on par with a WaveNet vocoder. 

\end{abstract}
\section{Introduction}
\label{sec:intro}

Recent advances in deep learning have led to text-to-speech (TTS) systems achieving near-human naturalness \cite{Shen2018-tacotron2}. 
This is partially due to neural sequence-to-sequence mapping methods that can learn to align and map between input text and output acoustic feature sequences 
\cite{Sotelo2017char2wav}. 
Another major progress in TTS is the introduction of waveform generation methods, such as WaveNet \cite{oord2016-wavenet}, that have been adopted to use as ``neural vocoders'' \cite{Tamamori2017-wavenet-vocoder}.
Generally, a neural vocoder is a neural network that, given some input acoustic features, outputs (speech) waveforms. This approach effectively decouples the acoustic mapping and waveform generation models from each other. Indeed, WaveNet-based neural vocoders have been successfully applied in TTS using either sequence-to-sequence techniques \cite{Shen2018-tacotron2} or more traditional statistical parametric speech synthesis (SPSS) acoustic models
\cite{Wang2018-comparison-of-waveform-generation}.
Although WaveNets and other autoregressive models produce impressive results, their inference process is inherently slow, and requires heavy optimization for real time applications \cite{kalbrenner2018-wave-rnn, Arik2017-deepvoice}.
Alternative approaches capable of parallel inference have been proposed \cite{oord2017-parallel-wavenet, ping2018-clarinet-parallel-wave-generation}, but these models are increasingly complex and difficult to train. 

Meanwhile, generative adversarial networks (GANs) have achieved impressive results in image synthesis \cite{Karras2018-prog-grow-gans}, and they are currently of increasing interest in speech applications. 
Recent advances are powered in part by more appropriate architectures, including residual connections 
\cite{Mescheder2018-which-gan-training-methods-converge}
progressive upsampling \cite{Karras2018-prog-grow-gans, Wang2018-prog-super-resolution} and multi-scale processing \cite{Denton2015-LAPGAN}. Similar mechanisms can be seen in the WaveNet architecture (cf. exponentially growing dilations and residual connections).
On the other hand, improved training techniques (e.g.~\cite{Gulrajani2017-wgan-gp-arxiv,Mescheder2018-which-gan-training-methods-converge}) have also resulted in better stability and convergence properties for GANs.
Nevertheless, while some speech-related GAN applications apply direct waveform-to-waveform transformations \cite{Pascual2017-segan, tanaka2018-wave-cycle-gan}, most TTS applications have so far used GANs on improving acoustic models instead of directly generating waveforms \cite{Kaneko2017-gan-postfilter, zhao2018-wgan-tts-with-wavenet}. 

Another approach to facilitate neural waveform generation for speech synthesis is the pitch-synchronous utilization of glottal waveforms (i.e. signals representing the true air flow excitation of speech generated by the vocal folds). In this approach, glottal inverse filtering \cite{Alku2011} is applied to the speech signal in order to remove the vocal tract resonances and to obtain a more elementary signal that is easier to model.
Early approaches used a point-wise least squares loss in time-domain \cite{Raitio2014b-voice-source-modelling-using-deep-neural, juvela2018-glotnet-interspeech}, and while this captures the gross waveshape well, the produced output is essentially a conditional average and lacks in stochastic high frequency contents (due to the averaging). The missing stochastic component can be recreated using signal processing techniques for aperiodicity modification, resulting in high quality synthetic speech
\cite{Airaksinen2018-vocoder-comparison, Cui2018-neural-glottal-vocoder}, but this involves making signal model assumptions that may not hold generally.  Further efforts have been made to model the stochastic part directly using GANs \cite{Bollepalli2017-gan-glottal-excitation} or WaveNet (``GlotNet'') \cite{juvela2018-glotnet-interspeech}. However, the former approach suffers from the unstable GAN training techniques available at the time, and for the latter  WaveNet inference limitations still apply. %
%
A similar excitation modeling approach can be extended to generating residual excitation signals for waveform synthesis from MFCCs, as MFCCs can be readily interpreted as spectral envelopes and cancelled from the signal via inverse filtering \cite{juvela2018-synthesis-from-mfcc}.

This paper proposes a novel multi-scale GAN architecture for pitch-synchronous waveform generation. The proposed generator performs progressive upsampling of feature maps and outputs waveforms at multiple timescales, while the discriminator ensures that the waveforms remain valid at each timescale. The model is trained using a Wasserstein GAN \cite{Gulrajani2017-wgan-gp-arxiv} with modified gradient penalties \cite{Mescheder2018-which-gan-training-methods-converge} and an FFT-based auxiliary loss, leading to  stable training behavior for both direct speech waveform and glottal excitation signals. The proposed model is evaluated as a neural vocoder for an SPSS system and compared with the GlottDNN and WaveNet vocoders. The results show that the proposed ``GlotGAN'' can achieve similar performance to the WaveNet vocoder, while the direct waveform ``WaveGAN'' still falls short in performance compared to the other methods. Computational benefits of the proposed model include parallel inference (due to GAN) and explicit fundamental frequency control (due to pitch-synchronous processing).







\section{Speech synthesis system}

The focus of this paper is on neural vocoders, and we use a conventional SPSS pipeline for the text-to-acoustic-features mapping. 
First, text is converted to linguistic features using the Flite speech synthesis front-end \cite{black2001flite} and the Combilex lexicon \cite{richmond2009combilex}. Alignments between the linguistic and acoustic features are found using the HMM-based speech synthesis system (HTS) \cite{Zen2007} and we use a bidirectional long short-term memory (BLSTM) recurrent neural network (RNN) for the acoustic model.
For system details, see \cite{juvela2017a-reduce-mismatch}.

This paper uses a common acoustic feature set of glottal vocoder features \cite{Airaksinen2018-vocoder-comparison} for all neural vocoders: 30 vocal tract filter line spectral frequencies (LSFs), 10 glottal source spectral envelope LSFs, 5 harmonic-to-noise ratio (HNR) parameters, fundamental frequency value on mel-scale (interpolated over unvoiced frames) and a binary voicing flag. 

\subsection{Speech material}

In the experiments, we use speech data from two speakers
(one male, one female), who both are professional British English voice talents. The dataset for the male speaker ``Nick'' comprises  2\,542 utterances, totaling 1.8 hours, and the dataset for the female speaker ``Jenny'' comprises 4\,314 utterances, totaling 4.8 hours.
For both speakers, 100-utterance test and validation sets were randomly selected from the data, while the remaining utterances were used for training. The material is used at a 16\,kHz sample rate.

\section{Proposed waveform generation model}


\subsection{Waveform representation}

\begin{figure*}[thb]
    \centering
    \subfloat{ \includegraphics[width=0.45\linewidth,trim={2cm 1.5cm 1cm 0.5cm},clip]{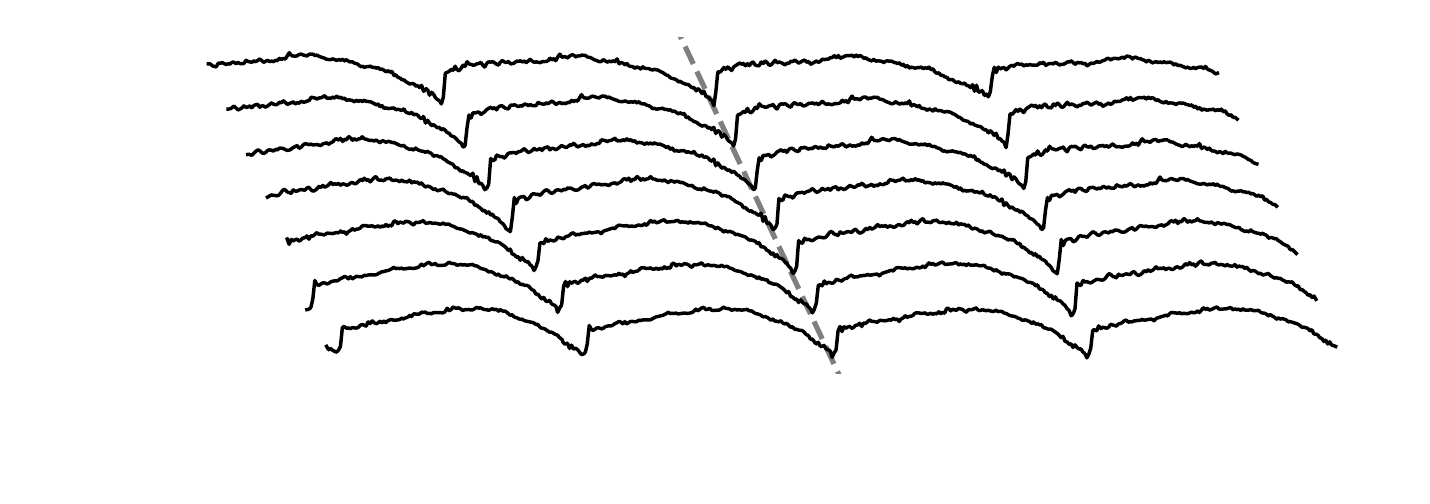}}%
\hfil
\subfloat{\includegraphics[width=0.45\linewidth,trim={2cm 1.5cm 1cm 0.5cm},clip]{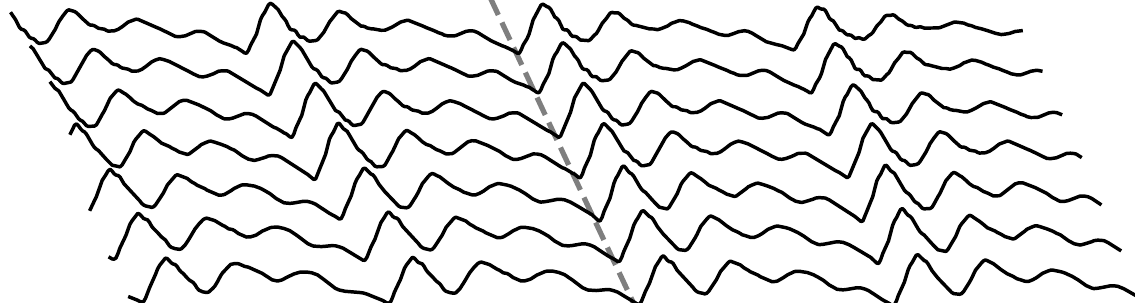}}%
    \caption{A waterfall plot showing consecutive frames of GCI-centered glottal excitation (left) and speech (right) waveforms.}
    \label{fig:waveform representations}
\end{figure*}

In this work, we further simplify the waveform representation from 
\cite{juvela2016a-high-pitched-excitation} by removing the pitch-adaptive cosine windowing and merely phase-lock the window midpoint to a glottal closure instant (GCI). See Fig. \ref{fig:waveform representations} for illustration. We use REAPER for GCI and pitch estimation \cite{talkin2015-reaper}. Similar phase-locked representations have been successfully applied not only in our previous work (\cite{juvela2016a-high-pitched-excitation,Bollepalli2017-gan-glottal-excitation, juvela2018-synthesis-from-mfcc}), but also in  \cite{Cui2018-neural-glottal-vocoder, Hwang2018-glottal-pulses-unified}. 
Furthermore, using GCIs to center waveforms in a window can be seen analogous to 
using facial landmarks to center images, as done in the highly successful CelebA-HQ dataset \cite{Karras2018-prog-grow-gans}. 

The target waveform can be the glottal excitation or the speech waveform directly. Our primary interest in this paper is the glottal excitation, as it appears relatively simpler to model \cite{juvela2018-glotnet-interspeech}, but we also include experiments on modeling the speech waveform directly. 
At synthesis time, the generated waveforms are assembled using pitch-synchronous overlap-add (PSOLA) \cite{moulines1995-psola}, but in principle, the non-tapered windowing enables using other concatenation techniques, including waveform similarity overlap-add (WSOLA) \cite{Verhelst1993-wsola}. When modeling glottal excitation signals, the assembled excitation signal is further filtered with the vocal tract filter to produce speech.


\subsection{Model architecture}

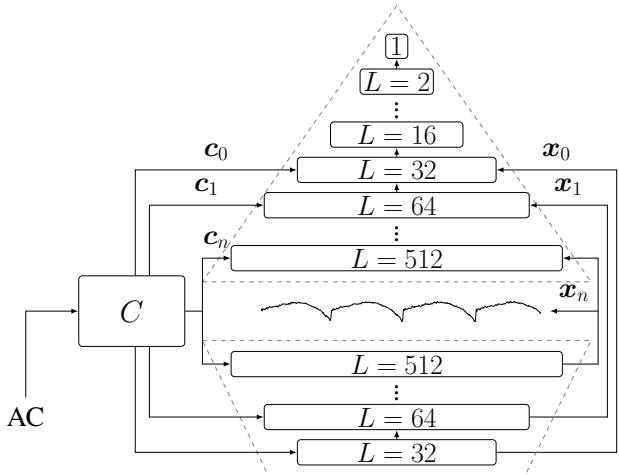
\begin{figure}[!t]
\centering
{
\fontsize{18}{20}\selectfont
\resizebox{0.99\linewidth}{!}{\input{figures/gan_architecture.tex}}%
}
\caption{ GAN architecture overview. Generator (bottom) performs progressive upsampling and outputs waveforms at multiple timescales, while the discriminator (top)  evaluates the waveforms at all timescales.  Both models have access to the same acoustic conditioning, provided by a conditioning model $C$ which collaborates with the generator.}
\label{fig:architecture_overview}
\end{figure}

A general view of our GAN architecture is shown in Fig.~\ref{fig:architecture_overview}.
Progressive upsampling aims for the lower resolution layers to learn the low frequency global structure, while the high resolution layers can focus on the high frequency stochastic signal components. By construction, the model allocates more capacity on the perceptually more relevant low frequencies.

\begin{figure}[thb]
\centering
{
\fontsize{22}{24}\selectfont
\resizebox{0.70\linewidth}{!}{\input{figures/generator_block.tex}}%
}
\caption{Generator upsampling block. Input hidden features $\bm h_{i-1}$ are upsampled linearly before concatenating them with a latent noise input $z_u$ and conditioning $c_i$. Each block applies a gated convolution to the concatenated features and outputs a signal $\bm x_i$ at its respective timescale}
\label{fig:generator_module}
\end{figure}
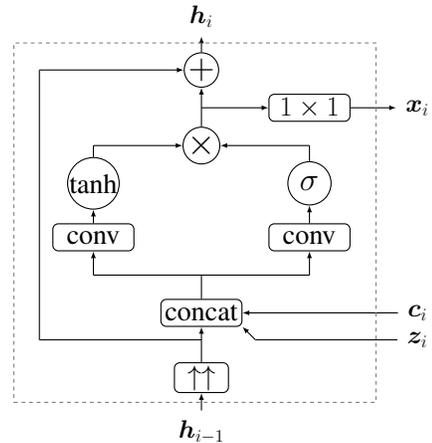

The generator $G$ consists of gated convolutional residual blocks 
with progressive feature map upsampling at each block. Fig.~\ref{fig:generator_module} shows a schematic of the generator block. 
Upsampling is performed along the sample dimension using linear interpolation to avoid checkerboard artefacts commonly arising from transposed convolution upsampling \cite{odena2016-deconvolution-checkerboard}.  In total, the generator contains five blocks that perform progressive upsampling from 32 points to 512 points. 

The discriminator $D$ consists of similar gated convolution blocks (without residual connections). Each block downsamples the input by a factor of two, using strided convolutions, until input width of one is reached.  The first five discriminator blocks concatenate the generator output or real image and the conditioning to the hidden activations at their respective timescale before applying the convolution.
Both the generator and the discriminator use 128-channel 2D-convolutions with filter width three across frames and seven across samples.

The conditioning model $C$ encodes the 200\,Hz frame rate acoustic sequence with a non-causal WaveNet-like dilated convolution structure (similar to \cite{Rethage2018-wavenet-speech-denoising}). The model consists of 
eight residual blocks with the dilation pattern $\lbrace1,2,4,8\rbrace$ repeated twice. The dilation stack is attached via skip connections to a projection module that outputs conditioning at each time-scale. The residual blocks use 64 channels and a filter width three.

\subsection{GAN training}


Denote the full-resolution signal segment by $\bm x_n$, and a collection $\bm x $ containing the segment $\bm x_n$ at all time resolutions by 
$\bm x = \lbrace \bm{x}_i \rbrace_{i = 0, \ldots, n} $, 
i.e. $\bm x_i$ is $\bm x_n$ downsampled by a factor of $2^i$. Similarly, the generator model $G$ outputs a collection of synthetic data points $\hat{\bm x} = G(\bm z, \bm c)$  at all timescales, given a Gaussian noise input $\bm z$ and time-varying conditioning $\bm c$ (derived from the acoustic feature input).

The goal of the generator is to produce $\hat{\bm x}$ that appears correct at each timescale, while the discriminator provides useful learning signals to match the real and generated data distributions (also at each timescale).  
We use the Wasserstein GAN \cite{Gulrajani2017-wgan-gp-arxiv} loss 
\begin{equation}
    \L^\mathrm{W}_D = -\E_{x \sim p_\mathcal{D} } \left[ D(\bm x, \bm c) \right] +  \E_{ \hat{x} \sim p_g} \left[ D( \hat{\bm x} , \bm c ) \right]
\end{equation}
for the discriminator and $\L^\mathrm{W}_G =  -\L^\mathrm{W}_D$ for the generator.
To keep the discriminator Lipschitz continuous, we use a one-sided gradient penalty (also proposed in \cite{Gulrajani2017-wgan-gp-arxiv}, see their Appendix C). 
We prevent the penalty from activating for magnitudes below one to avoid cyclic, non-convergent training dynamics described in \cite{Mescheder2018-which-gan-training-methods-converge}. The masked gradient penalty is given by
\begin{equation}
    \L^\mathrm{GP}_D = \E_{x \sim p_\mathcal{D}, \hat{x} \sim p_g } \left[ \left( \max \lbrace 0,  \norm{ \nabla_{\tilde{\bm x}} D(\tilde{\bm x}, \bm c )} - 1 \rbrace \right)^2 \right],
\end{equation}
where $\tilde{\bm x} = \varepsilon \bm x + (1-\varepsilon) \hat{\bm x}$ is sampled randomly along the line segment between $\bm x$ and $\hat{\bm x}$.
Additionally, to encourage convergence, we use the ``R1'' penalty \cite{Mescheder2018-which-gan-training-methods-converge} that penalizes large discriminator gradient magnitudes for the real data samples
\begin{equation}
    \L^\mathrm{R1}_D = \E_{x \sim p_\mathcal{D}} \left[  \norm{\nabla_{\bm x} D(\bm x, \bm c )}^2 \right]
\end{equation}

Finally, we add an optional loss term based on the mean squared error of FFT magnitudes. Similar auxiliary FFT-based losses have been found helpful for waveform generation in e.g. \cite{oord2017-parallel-wavenet,juvela2018-synthesis-from-mfcc,ping2018-clarinet-parallel-wave-generation}. We only apply the FFT loss on the full-resolution signal:
\begin{equation}
    \L^\mathrm{FFT}_G = \E \left[ \left( \lvert \mathrm{FFT}(\bm x_n) \rvert -  \lvert \mathrm{FFT}(\hat{\bm x}_n) \rvert \right)^2 \right]
\end{equation}
Notably, the FFT-based loss only uses the spectral magnitude, so all learning of phase information is due to the GAN loss. 
We found that the spectral loss stabilizes the training both for speech waveforms and glottal excitations, but is not strictly necessary for the latter.

The total training objective is to minimize $\L_{G,C} = \L^\mathrm{W}_G + \lambda_1 \L^\mathrm{FFT}_G$ by updating the generator and the conditioning model, while minimizing $\L_D = \L^\mathrm{W}_D + \lambda_2 \L^\mathrm{GP}_D +  \lambda_3 \L^\mathrm{R1}_D $ by updating the discriminator. 
We use alternating gradient descent with Adam 
(LR=1e-4, $\beta_1$=0.9, $\beta_2$=0.999) and loss weights $\lambda_1$=1, $\lambda_2$=10, $\lambda_3$=1. Similarly to \cite{Karras2018-prog-grow-gans}, we only apply one discriminator update per generator update and use an additional batch standard deviation feature in the penultimate discriminator layer. The models were trained for 100k iterations, where a single iteration contains 150 consecutive frames of speech, each associated with a 512 point full resolution waveform.

\section{Evaluation}

We conducted subjective listening tests to evaluate the synthetic speech quality and voice similarity to a natural reference. Our main interest is the comparison between the proposed method, applied to glottal excitation signals (named ``GlotGAN''), and two established neural vocoder methods, GlottDNN and WaveNet. For the similarity DMOS test, we included a direct waveform variant of the proposed method (called here ``WaveGAN'') and a classical SPSS vocoder, STRAIGHT \cite{Kawahara1999-straight-envelope} (which uses its distinct feature set and acoustic model). The latter two were excluded from the pair-wise quality CCR test, as the number of system pairings would have grown to be unpractical.
Audio samples and code are available.\footnote{\url{https://github.com/ljuvela/multiscale-GAN}}

\subsection{Reference methods}


The GlottDNN and Wavenet vocoders are both conditioned on the same acoustic feature set as the proposed model. GlottDNN uses a DNN to predict glottal excitation pulse waveshapes in a two period pitch synchronous format \cite{juvela2016a-high-pitched-excitation}. The excitation model outputs conditional average pulse shapes, where the lack of high frequency stochastic content is compensated by adding shaped noise as indicated by the HNRs, and by applying a spectral envelope matching filter \cite{Airaksinen2018-vocoder-comparison}.
We use the configuration from \cite{juvela2017a-reduce-mismatch}, where the excitation model consists of a single BLSTM layer (size 128), followed by three fully connected layers (size 512), finally outputting a 400 point pulse.  
Our WaveNet vocoder uses a model configuration from \cite{juvela2018-glotnet-interspeech} (30 residual blocks in three dilation groups, 64 residual channels, 128 post-net channels). However, the model is trained with 8-bit softmax cross-entropy on $\mu$-law companded quantized speech, and is trained on the same speaker dependent data as the proposed GlotGAN model. 

\subsection{Listening test}

Listening tests were conducted on the Figure Eight\footnote{\url{https://www.figure-eight.com/}} 
crowd-sourcing platform.  Each test case was evaluated by 50 listeners, and listener quality was maintained by artificial low-quality anchor cases and post-screening of zero-variance listener responses. 15 test set utterances were randomly selected for the listening experiments.

\begin{figure}[thb]
\includegraphics[height=0.65\linewidth]{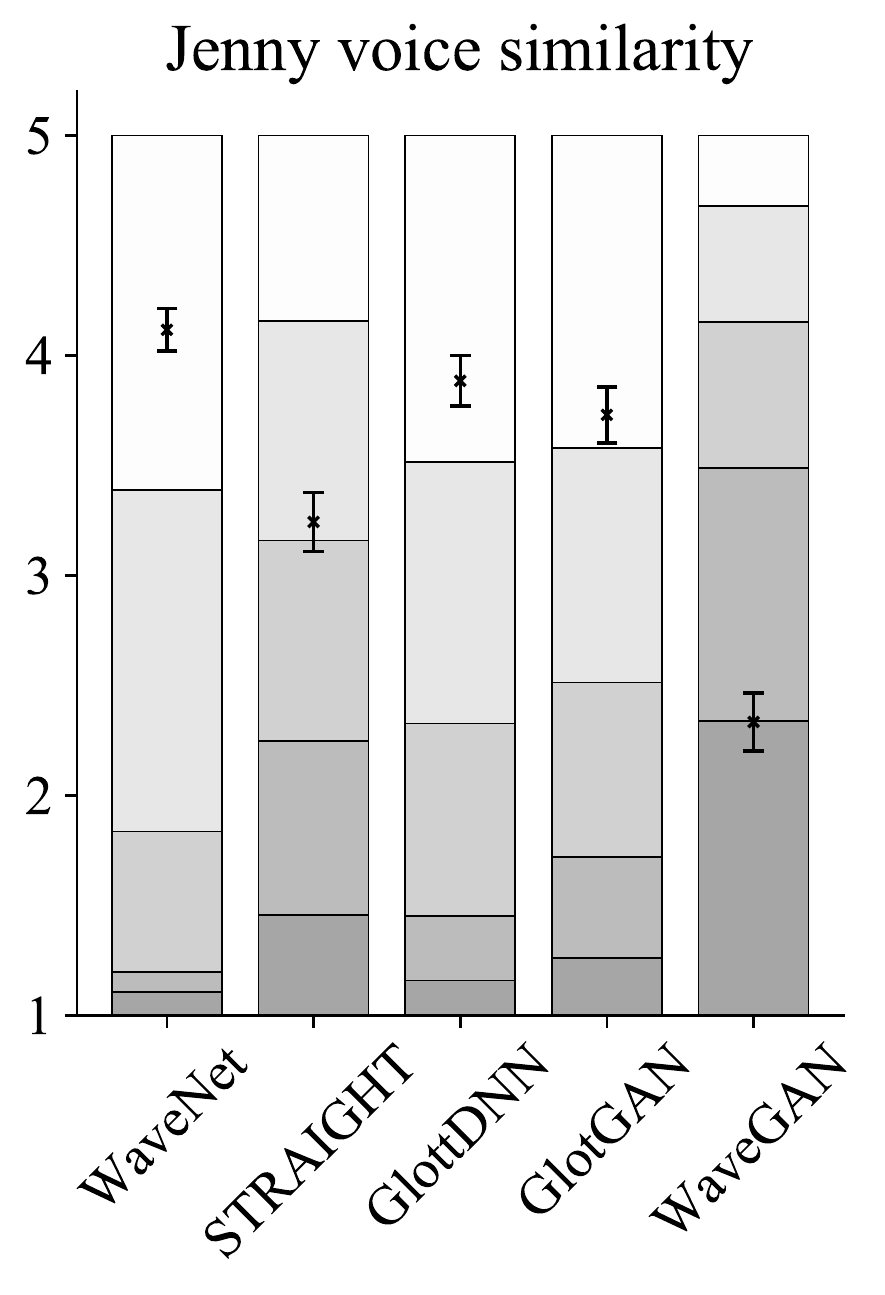}
\includegraphics[height=0.65\linewidth]{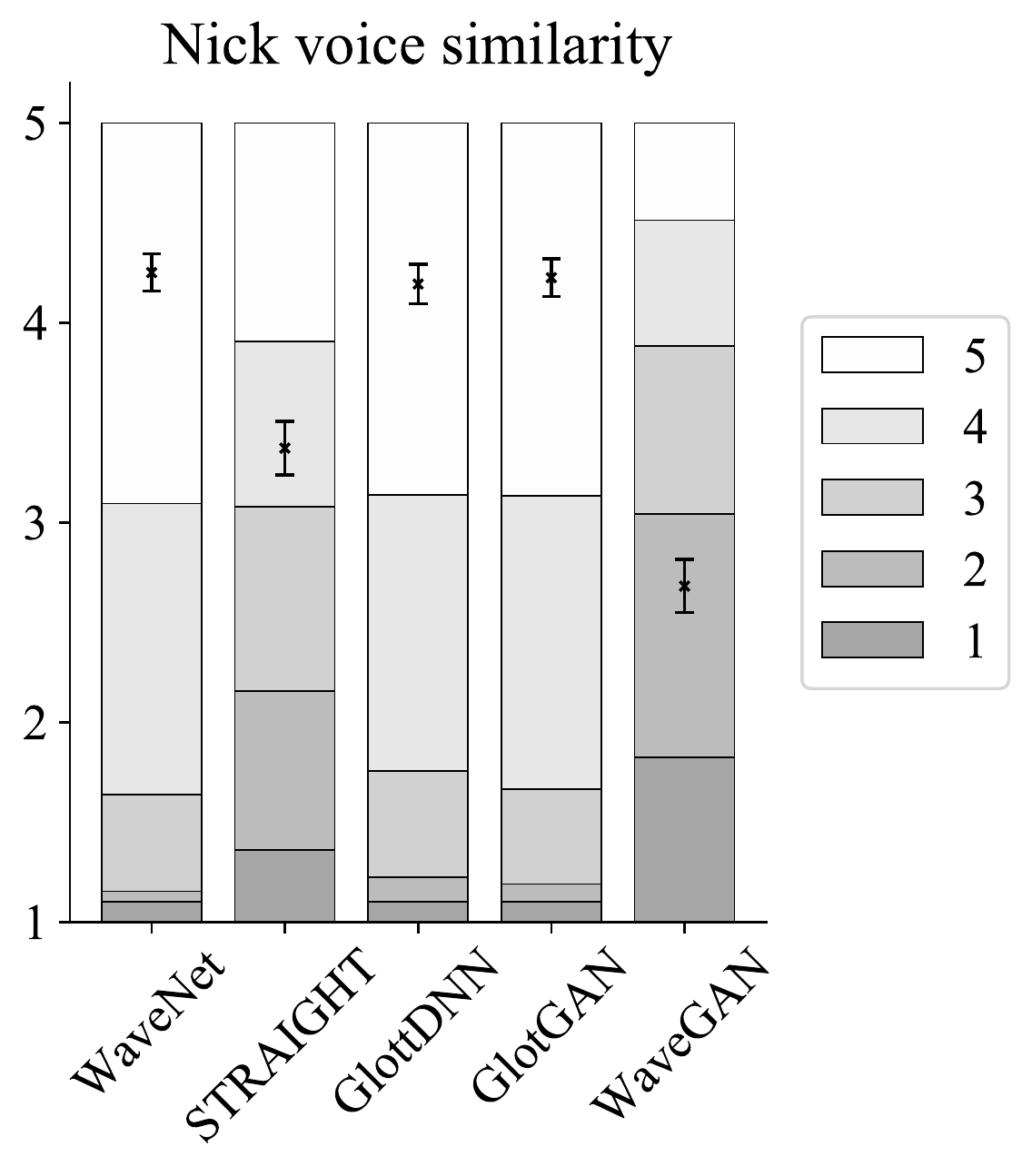}
\caption{Voice similarity ratings for ``Jenny'' (left) and ``Nick'' (right). The plot shows mean ratings with 95\% confidence intervals, along with stacked rating distribution histograms.}
\label{fig:DMOS_voice_sim}
\end{figure}

Voice similarity between synthetic speech and a natural reference was evaluated in a DMOS-like test. 
The listeners were asked to rate the voice similarity of a test sample to a natural reference (with the same linguistic contents) using a 5-level absolute category rating scale, ranging from ``Bad''(1) to ``Excellent''(5).   
Figure \ref{fig:DMOS_voice_sim} shows system mean ratings with 95\% confidence intervals and normalized histograms for the rating categories. The Mann-Whitney
 U-tests indicate that differences between systems are statistically significant (at a Bonferroni corrected 95\% confidence level), except GlotGAN--GlottDNN for ``Jenny'' and all pairings of WaveNet--GlotGAN--GlottDNN for ``Nick''.

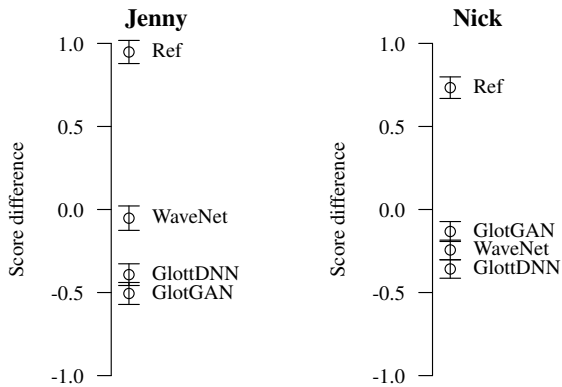
\begin{figure}[thb]
\centering
\subfloat{\resizebox{0.49\linewidth}{!}{\input{figures/ccr_jenny.tex}}}%
\hfil
\subfloat{\resizebox{0.49\linewidth}{!}{\input{figures/ccr_nick.tex}}}%
\caption{Combined score differences obtained from the quality comparison CCR test for ``Jenny'' (left) and ``Nick'' (right). Error bars are t-statistic based 95\% confidence intervals for the mean.}
\label{fig:CCR_quality}
\end{figure}

A category comparison rating (CCR) \cite{Itu1996} test was performed to evaluate the synthetic speech quality. The listeners were presented with a pair of test samples and asked to rate the comparative quality from -3 (``Much worse'') to 3 (``Much better''). The CCR scores are obtained by re-ordering and pooling together all ratings the system received. Figure \ref{fig:CCR_quality} shows the mean scores with 95\% confidence intervals. U-tests (following a similar procedure to above) indicate that all differences are statistically significant. 

\section{Discussion}

A characteristic issue we have identified with frame-based GAN audio generation is  related to local GAN mode collapse (i.e.~lack of variety). In this case, the Generator network fails to reproduce the correct stochastic properties of the data, and instead outputs a ``mode'' waveshape that appears appropriate when examined in isolation, but causes audible artefacts when combined with adjacent frames.
Perceptually, this repeated ``frozen noise'' can result in buzzy robotic artefacts and ringing, reminiscent of pure impulse train excited vocoders. 
This issue is especially prominent in unvoiced sounds, and for the listening experiments we resorted to synthesizing the unvoiced parts using pure white noise excitation filtered with the predicted vocal tract envelope. This effect is also somewhat audible in the voiced parts of GlotGAN for ``Jenny'', which contributes to GlotGAN under-performing the reference methods. As for direct waveform synthesis, the quality is further degraded, although the synthetic speech remains intelligible.

Another potential cause to these artefacts is overfitting and the mismatch between natural and synthetic acoustic features at training and test time, respectively. Furthermore, the acoustic model performance varies between different speakers \cite{Airaksinen2018-vocoder-comparison}, as reflected in the larger gap between natural and synthetic voices for ``Jenny'' compared with ``Nick''. In contrast, the difference in neural vocoder performance in less prominent with the same speech material, when using natural acoustic features in a copy-synthesis setup \cite{juvela2018-synthesis-from-mfcc}.

\section{Conclusion}
This paper presented a multi-scale GAN architecture for generating glottal excitation (GlotGAN) and speech waveforms (WaveGAN) pitch-synchronously. The proposed model is evaluated as a neural vocoder for a statistical parametric speech synthesis system and listening test results show that GlotGAN can achieve similar performance to a WaveNet vocoder.
Future work includes upgrading the TTS system acoustic mapping to use sequence-to-sequence models \cite{Shen2018-tacotron2} and applying joint generative acoustic model and neural vocoder training \cite{zhao2018-wgan-tts-with-wavenet}. 
Progressive upsampling should also naturally extend to higher sample rates, while still allowing fast parallel inference.

\section{Acknowledgment}
This study was supported by the Academy of Finland (project 312490). JY was partially supported by JST CREST Grant Number JPMJCR18A6, Japan and by MEXT KAKENHI Grant Numbers (16H06302, 17H04687, 18H04120, 18H04112, 18KT0051), Japan.
We acknowledge the computational resources provided by the Aalto Science-IT project.


{
\bibliographystyle{IEEEbib}
\bibliography{refs}
}

\end{document}

%% file: figures/gan_architecture.tex
\def\BlockWidth{0.50cm}
\def\Unit{0.8cm}

\begin{tikzpicture}
    [align=center, node distance=\Unit, line width=0.2mm]

    \tikzstyle{rectnode}=[rectangle, rounded corners={0.01*\Unit}, inner sep=2pt, minimum width={3*\Unit}, minimum height={\BlockWidth}, draw, fill=white, rounded corners={0.1*\Unit}]
    \tikzstyle{line}=[-latex]

    \coordinate (zero) at (0.0, 0.0); 

    \node[rectnode, below of = zero, node distance = {1.5*\Unit}, minimum width = {15*\BlockWidth}]
         (G_0)  {$L=512$};
    \node[rectnode, below of = G_0, minimum width = {12*\BlockWidth}, node distance = {1.5*\Unit}] 
        (G_1) {$L=64$};
    \node[rectnode, below of = G_1, minimum width =  {9*\BlockWidth}]
        (G_2) {$L=32$};

    \node[rectnode, above of = zero, node distance = {1.5*\Unit}, minimum width = {15*\BlockWidth}]
        (D_0) {$L=512$};
    \node[rectnode, above of = D_0, minimum width = {12*\BlockWidth},  node distance = {1.5*\Unit}]
        (D_1) {$L=64$};
    \node[rectnode, above of = D_1, minimum width = {9*\BlockWidth}]
        (D_2) {$L=32$};
    \node[rectnode, above of = D_2, minimum width = {6*\BlockWidth}]
        (D_3) {$L=16$};
    \node[rectnode, above of = D_3, minimum width = {3*\BlockWidth}, node distance = {1.5*\Unit}]
        (D_4) {$L=2$};
    \node[rectnode, above of = D_4, minimum width = {1*\BlockWidth}] 
        (D_5) {$1$};

    \node[] () {$G$};

    \draw[line] (G_2) -- (G_1);
    \node[](G_dots) at ($(G_1)!0.55!(G_0)$) {$\vdots$};
    \node[] (D_dots) at ($(D_1)!0.45!(D_0)$) {$\vdots$};
    \draw[line] (D_1) -- (D_2);
    \draw[line] (D_2) -- (D_3);
    \node[] (D_dots2) at ($(D_3)!0.55!(D_4)$) {$\vdots$};
    \draw[line] (D_4) -- (D_5);
    
    \coordinate (output_0) at ($ (G_0.east |- 0,0 ) + (1.0*\Unit, 0)$) ;
    \draw[line] (G_0.east) --  (G_0.east -| output_0) -- (D_0.east -| output_0) -- (D_0.east);
    \coordinate (output_1) at ($ (G_1.east |- 0,0 ) + (2.2*\Unit, 0)$) ;
    \draw[line] (G_1.east) -- (G_1.east -| output_1) -- (D_1.east -| output_1) -- node[above] {$\bm x_1$}  (D_1.east);
    \coordinate (output_2) at ($ (G_2.east |- 0,0 ) + (3.4*\Unit, 0)$) ;
    \draw[line] (G_2.east) -- (G_2.east -| output_2) -- (D_2.east -| output_2) -- node[above] {$\bm x_0$}  (D_2.east);

    \node (waveform) at (0,0) {\includegraphics[width=6.5cm, trim={2.3cm 1.3cm 1cm 0.8cm}, clip]{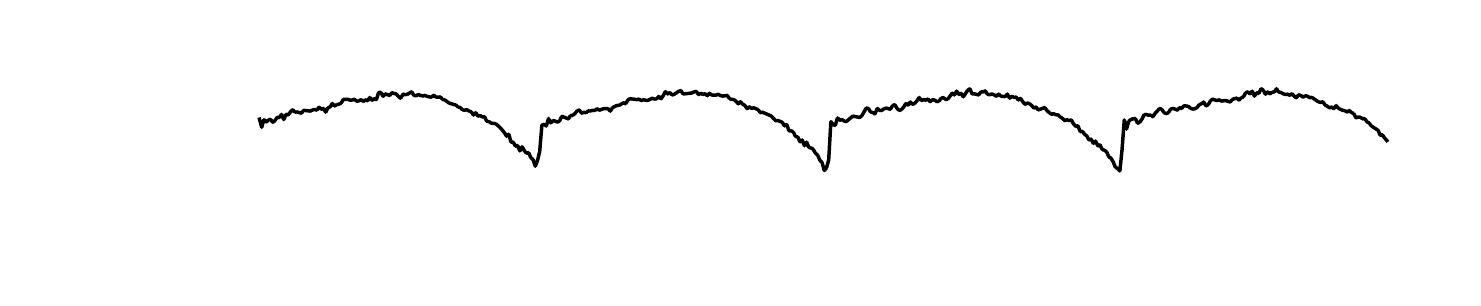}};
    \draw[line] (output_0) --  node[above] {$\bm x_n$} (waveform.east);  


    \draw[color=black!50, opacity=1.0, dashed]
        ($(G_0.north west) + (-0.8*\Unit, 0.3*\Unit)$) -- 
        ($(G_0.north east) + (0.8*\Unit, 0.3*\Unit)$) --
        ($(G_2.south east) + (0.8*\Unit, -0.3*\Unit)$) --
        ($(G_2.south west) + (-0.8*\Unit, -0.3*\Unit)$) --  
        ($(G_0.north west) + (-0.8*\Unit, 0.3*\Unit)$)  node[above, color=black] {};

    \draw[color=black!50, opacity=1.0, dashed]
        ($(D_0.south west) + (-0.8*\Unit, -0.3*\Unit)$) -- 
        ($(D_0.south east) + (0.8*\Unit, -0.3*\Unit)$) --
        ($(D_5.north) + (0, 0.8*\Unit)$) -- node[above, color=black, node distance= 2.0cm] {}
        ($(D_0.south west) + (-0.8*\Unit, -0.3*\Unit)$)  ;

    \node[rectnode, left of = zero, node distance = {7.5*\Unit}, minimum height={2.0*\Unit}]
        (cond) {\fontsize{0.7cm}{0.7cm}\selectfont$C$};    

    \coordinate[right of=cond, node distance = {2*\Unit}] (cond_split_0); 
    \draw[line] (cond) -- (cond_split_0) -- (cond_split_0 |- G_0)  -- (G_0);
    \draw[line] (cond_split_0) -- (cond_split_0 |- D_0) --  node[above, midway] {$\bm c_n$}  (D_0);

    \coordinate[right of=cond, node distance = {0.5*\Unit}] (cond_split_1); 
    \draw[line] (cond_split_1 |- cond.north) -- (cond_split_1 |- D_1) --  node[above, midway] {$\bm c_1$}  (D_1);
    \draw[line] (cond_split_1 |- cond.south) -- (cond_split_1 |- G_1) -- (G_1);

    \coordinate[right of=cond, node distance = {0.1*\Unit}] (cond_split_2); 
    \draw[line] (cond_split_2 |- cond.north) -- (cond_split_2 |- D_2) --  node[above, midway] {$\bm c_0$}  (D_2);
    \draw[line] (cond_split_2 |- cond.south) -- (cond_split_2 |- G_2) -- (G_2);    


    \coordinate[left of = cond, node distance = {3*\Unit}](left_of_cond);
    \node[below of = left_of_cond, node distance = {3*\Unit}] (ac_input) {AC};
    \draw[line] (ac_input) -- (left_of_cond) -- (cond);

\end{tikzpicture}

%% file: figures/generator_block.tex
\def\Unit{1.5cm}
\def\LineWidth{0.3mm}
\def\MathFontSize{\fontsize{0.7cm}{0.7cm}\selectfont}

\begin{tikzpicture}
    [align=center,node distance={0.7*\Unit}, line width= \LineWidth]

    \tikzstyle{rectnode}=[rectangle, inner sep=2pt, minimum width={1.5*\Unit}, minimum height={0.5*\Unit}, draw, fill=white, rounded corners={0.1*\Unit}]
    \tikzstyle{roundnode}=[circle, inner sep=0pt, minimum width={0.8*\Unit}, minimum height={0.1*\Unit}, draw, fill=white]

    \tikzstyle{line}=[-latex, line width=\LineWidth]

    \node (input) at (0.0, 0.0) {\MathFontSize $\bm h_{i-1}$};
    \node [rectnode, minimum width = {1.0*\Unit}, above of = input, node distance = {1.0*\Unit}] (upsample) { $\uparrow \uparrow$};
    \coordinate [above of = upsample] (input_split1);
    \node [rectnode, node distance = {0.5*\Unit}, above of = input_split1] (concat) {concat};
    \coordinate [above of = concat] (input_split2);
    \draw[line] (input) -- (upsample);
    \draw[line] (upsample) -- (concat);
    \draw[] (concat)  -- (input_split2);

    \coordinate [right of = input_split2, node distance = {2*\Unit}] (right_bottom);
    \coordinate [left of = input_split2, node distance = {2*\Unit}] (left_bottom);
    \draw[] (input_split2) -- (right_bottom);
    \draw[] (input_split2) -- (left_bottom);

    \node[rectnode, above of = left_bottom] (left_conv) {conv};
    \draw[line] (left_bottom) -- (left_conv);
    \node[rectnode, above of = right_bottom] (right_conv) {conv};
    \draw[line] (right_bottom) -- (right_conv);
    
    \node[roundnode, above of = left_conv, node distance = {1.0*\Unit}] (tanh) {tanh};
    \draw[line] (left_conv) -- (tanh) ;
    \node[roundnode, above of = right_conv, node distance = {1.0*\Unit}] (sigma) {\fontsize{0.9cm}{0.9cm}\selectfont $\sigma$};
    \draw[line] (right_conv) -- (sigma);

    \coordinate[above of = tanh] (left_top);
    \draw[] (tanh) -- (left_top);

    \coordinate[above of = sigma] (right_top);
    \draw[] (sigma) -- (right_top);

    \node[roundnode, minimum width = {0.5*\Unit}] (multiply) at (right_top -| input) {\fontsize{1cm}{1.1cm}\selectfont $\times$}; 

    \draw[line] (left_top) -- (multiply);
    \draw[line] (right_top) -- (multiply);

    \coordinate[above of = multiply] (skip_split) ; 

    \node[rectnode, right of = skip_split, node distance = {2*\Unit}] (skip_conv) {$1 \times 1$}; 
    \draw[line] (skip_split) -- (skip_conv);
    \node[right of = skip_conv, node distance = {2*\Unit}] (skip_out) {\MathFontSize$\bm x_i$};
    \draw[line] (skip_conv) -- (skip_out);

    \node[] (cond_in) at (concat -| skip_out) {\MathFontSize$\bm c_i$};
    \draw[line] (cond_in) -- (concat);

    \coordinate[right of = input_split1, node distance = {1*\Unit}] (noise_anchor);
    \node[] (noise) at (input_split1 -| skip_out) {\MathFontSize$\bm z_i$};
    \draw[] (noise) -- (noise_anchor); 
    \draw[line] (noise_anchor) -- (concat.south east);

    \node[roundnode, minimum width = {0.5*\Unit}, above of = skip_split] (add) {\fontsize{1cm}{1.1cm}\selectfont $+$}; 
    \draw[line] (multiply) -- (add);

    \node[above of = add, node distance = {1.0*\Unit}] (output) {\MathFontSize $\bm h_i$};
    \draw[line] (add) -- (output);

    \coordinate[left of = input_split1, node distance = {3.0*\Unit}] (residual_bottom);
    \draw[] (input_split1) -- (residual_bottom);
    \coordinate[] (residual_top) at (residual_bottom |- add);
    \draw[] (residual_bottom) -- (residual_top);
    \draw[line] (residual_top) -- (add);

    \node [draw=black!50, dashed, fit={(upsample) (add)  ($(skip_conv.east) + (0.3*\Unit, 0)$) ($(residual_top) + (-0.3*\Unit, 0)$) }] {};

\end{tikzpicture}

%% file: figures/ccr_jenny.tex
\begin{tikzpicture}[x=1pt,y=1pt]
\definecolor[named]{fillColor}{rgb}{1.00,1.00,1.00}
\path[use as bounding box,fill=fillColor,fill opacity=0.00] (0,0) rectangle (108.41,144.54);
\begin{scope}
\path[clip] (  0.00,  0.00) rectangle (108.41,144.54);
\definecolor[named]{drawColor}{rgb}{0.00,0.00,0.00}

\path[draw=drawColor,line width= 0.4pt,line join=round,line cap=round] ( 54.19,127.30) circle (  1.80);

\path[draw=drawColor,line width= 0.4pt,line join=round,line cap=round] ( 54.19, 42.93) circle (  1.80);

\path[draw=drawColor,line width= 0.4pt,line join=round,line cap=round] ( 54.19, 69.24) circle (  1.80);

\path[draw=drawColor,line width= 0.4pt,line join=round,line cap=round] ( 54.19, 49.51) circle (  1.80);

\path[draw=drawColor,line width= 0.4pt,line join=round,line cap=round] ( 48.00, 14.24) -- ( 48.00,130.30);

\path[draw=drawColor,line width= 0.4pt,line join=round,line cap=round] ( 48.00, 14.24) -- ( 43.20, 14.24);

\path[draw=drawColor,line width= 0.4pt,line join=round,line cap=round] ( 48.00, 43.26) -- ( 43.20, 43.26);

\path[draw=drawColor,line width= 0.4pt,line join=round,line cap=round] ( 48.00, 72.27) -- ( 43.20, 72.27);

\path[draw=drawColor,line width= 0.4pt,line join=round,line cap=round] ( 48.00,101.28) -- ( 43.20,101.28);

\path[draw=drawColor,line width= 0.4pt,line join=round,line cap=round] ( 48.00,130.30) -- ( 43.20,130.30);

\node[text=drawColor,anchor=base east,inner sep=0pt, outer sep=0pt, scale=  0.80] at ( 38.40, 11.49) {-1.0};

\node[text=drawColor,anchor=base east,inner sep=0pt, outer sep=0pt, scale=  0.80] at ( 38.40, 40.50) {-0.5};

\node[text=drawColor,anchor=base east,inner sep=0pt, outer sep=0pt, scale=  0.80] at ( 38.40, 69.52) {0.0};

\node[text=drawColor,anchor=base east,inner sep=0pt, outer sep=0pt, scale=  0.80] at ( 38.40, 98.53) {0.5};

\node[text=drawColor,anchor=base east,inner sep=0pt, outer sep=0pt, scale=  0.80] at ( 38.40,127.54) {1.0};

\node[text=drawColor,rotate= 90.00,anchor=base,inner sep=0pt, outer sep=0pt, scale=  0.80] at ( 17.28, 72.27) {Score difference};

\path[draw=drawColor,line width= 0.4pt,line join=round,line cap=round] ( 54.19,123.26) -- ( 54.19,131.34);

\path[draw=drawColor,line width= 0.4pt,line join=round,line cap=round] ( 50.57,123.26) --
	( 54.19,123.26) --
	( 57.80,123.26);

\path[draw=drawColor,line width= 0.4pt,line join=round,line cap=round] ( 57.80,131.34) --
	( 54.19,131.34) --
	( 50.57,131.34);

\path[draw=drawColor,line width= 0.4pt,line join=round,line cap=round] ( 54.19, 39.10) -- ( 54.19, 46.77);

\path[draw=drawColor,line width= 0.4pt,line join=round,line cap=round] ( 50.57, 39.10) --
	( 54.19, 39.10) --
	( 57.80, 39.10);

\path[draw=drawColor,line width= 0.4pt,line join=round,line cap=round] ( 57.80, 46.77) --
	( 54.19, 46.77) --
	( 50.57, 46.77);

\path[draw=drawColor,line width= 0.4pt,line join=round,line cap=round] ( 54.19, 64.98) -- ( 54.19, 73.51);

\path[draw=drawColor,line width= 0.4pt,line join=round,line cap=round] ( 50.57, 64.98) --
	( 54.19, 64.98) --
	( 57.80, 64.98);

\path[draw=drawColor,line width= 0.4pt,line join=round,line cap=round] ( 57.80, 73.51) --
	( 54.19, 73.51) --
	( 50.57, 73.51);

\path[draw=drawColor,line width= 0.4pt,line join=round,line cap=round] ( 54.19, 45.72) -- ( 54.19, 53.30);

\path[draw=drawColor,line width= 0.4pt,line join=round,line cap=round] ( 50.57, 45.72) --
	( 54.19, 45.72) --
	( 57.80, 45.72);

\path[draw=drawColor,line width= 0.4pt,line join=round,line cap=round] ( 57.80, 53.30) --
	( 54.19, 53.30) --
	( 50.57, 53.30);

\node[text=drawColor,anchor=base west,inner sep=0pt, outer sep=0pt, scale=  0.80] at ( 62.33,125.47) {Ref};

\node[text=drawColor,anchor=base west,inner sep=0pt, outer sep=0pt, scale=  0.80] at ( 62.33, 41.10) {GlotGAN};

\node[text=drawColor,anchor=base west,inner sep=0pt, outer sep=0pt, scale=  0.80] at ( 62.33, 67.40) {WaveNet};

\node[text=drawColor,anchor=base west,inner sep=0pt, outer sep=0pt, scale=  0.80] at ( 62.33, 47.67) {GlottDNN};

\node[text=drawColor,anchor=base,inner sep=0pt, outer sep=0pt, scale=  0.96] at ( 63.80,136.43) {\bfseries Jenny};
\end{scope}
\end{tikzpicture}

%% file: figures/ccr_nick.tex
\begin{tikzpicture}[x=1pt,y=1pt]
\definecolor[named]{fillColor}{rgb}{1.00,1.00,1.00}
\path[use as bounding box,fill=fillColor,fill opacity=0.00] (0,0) rectangle (108.41,144.54);
\begin{scope}
\path[clip] (  0.00,  0.00) rectangle (108.41,144.54);
\definecolor[named]{drawColor}{rgb}{0.00,0.00,0.00}

\path[draw=drawColor,line width= 0.4pt,line join=round,line cap=round] ( 54.19,114.84) circle (  1.80);

\path[draw=drawColor,line width= 0.4pt,line join=round,line cap=round] ( 54.19, 64.58) circle (  1.80);

\path[draw=drawColor,line width= 0.4pt,line join=round,line cap=round] ( 54.19, 58.14) circle (  1.80);

\path[draw=drawColor,line width= 0.4pt,line join=round,line cap=round] ( 54.19, 51.49) circle (  1.80);

\path[draw=drawColor,line width= 0.4pt,line join=round,line cap=round] ( 48.00, 14.24) -- ( 48.00,130.30);

\path[draw=drawColor,line width= 0.4pt,line join=round,line cap=round] ( 48.00, 14.24) -- ( 43.20, 14.24);

\path[draw=drawColor,line width= 0.4pt,line join=round,line cap=round] ( 48.00, 43.26) -- ( 43.20, 43.26);

\path[draw=drawColor,line width= 0.4pt,line join=round,line cap=round] ( 48.00, 72.27) -- ( 43.20, 72.27);

\path[draw=drawColor,line width= 0.4pt,line join=round,line cap=round] ( 48.00,101.28) -- ( 43.20,101.28);

\path[draw=drawColor,line width= 0.4pt,line join=round,line cap=round] ( 48.00,130.30) -- ( 43.20,130.30);

\node[text=drawColor,anchor=base east,inner sep=0pt, outer sep=0pt, scale=  0.80] at ( 38.40, 11.49) {-1.0};

\node[text=drawColor,anchor=base east,inner sep=0pt, outer sep=0pt, scale=  0.80] at ( 38.40, 40.50) {-0.5};

\node[text=drawColor,anchor=base east,inner sep=0pt, outer sep=0pt, scale=  0.80] at ( 38.40, 69.52) {0.0};

\node[text=drawColor,anchor=base east,inner sep=0pt, outer sep=0pt, scale=  0.80] at ( 38.40, 98.53) {0.5};

\node[text=drawColor,anchor=base east,inner sep=0pt, outer sep=0pt, scale=  0.80] at ( 38.40,127.54) {1.0};

\node[text=drawColor,rotate= 90.00,anchor=base,inner sep=0pt, outer sep=0pt, scale=  0.80] at ( 17.28, 72.27) {Score difference};

\path[draw=drawColor,line width= 0.4pt,line join=round,line cap=round] ( 54.19,111.07) -- ( 54.19,118.60);

\path[draw=drawColor,line width= 0.4pt,line join=round,line cap=round] ( 50.57,111.07) --
	( 54.19,111.07) --
	( 57.80,111.07);

\path[draw=drawColor,line width= 0.4pt,line join=round,line cap=round] ( 57.80,118.60) --
	( 54.19,118.60) --
	( 50.57,118.60);

\path[draw=drawColor,line width= 0.4pt,line join=round,line cap=round] ( 54.19, 61.10) -- ( 54.19, 68.06);

\path[draw=drawColor,line width= 0.4pt,line join=round,line cap=round] ( 50.57, 61.10) --
	( 54.19, 61.10) --
	( 57.80, 61.10);

\path[draw=drawColor,line width= 0.4pt,line join=round,line cap=round] ( 57.80, 68.06) --
	( 54.19, 68.06) --
	( 50.57, 68.06);

\path[draw=drawColor,line width= 0.4pt,line join=round,line cap=round] ( 54.19, 54.71) -- ( 54.19, 61.57);

\path[draw=drawColor,line width= 0.4pt,line join=round,line cap=round] ( 50.57, 54.71) --
	( 54.19, 54.71) --
	( 57.80, 54.71);

\path[draw=drawColor,line width= 0.4pt,line join=round,line cap=round] ( 57.80, 61.57) --
	( 54.19, 61.57) --
	( 50.57, 61.57);

\path[draw=drawColor,line width= 0.4pt,line join=round,line cap=round] ( 54.19, 48.26) -- ( 54.19, 54.71);

\path[draw=drawColor,line width= 0.4pt,line join=round,line cap=round] ( 50.57, 48.26) --
	( 54.19, 48.26) --
	( 57.80, 48.26);

\path[draw=drawColor,line width= 0.4pt,line join=round,line cap=round] ( 57.80, 54.71) --
	( 54.19, 54.71) --
	( 50.57, 54.71);

\node[text=drawColor,anchor=base west,inner sep=0pt, outer sep=0pt, scale=  0.80] at ( 62.33,113.00) {Ref};

\node[text=drawColor,anchor=base west,inner sep=0pt, outer sep=0pt, scale=  0.80] at ( 62.33, 62.74) {GlotGAN};

\node[text=drawColor,anchor=base west,inner sep=0pt, outer sep=0pt, scale=  0.80] at ( 62.33, 56.30) {WaveNet};

\node[text=drawColor,anchor=base west,inner sep=0pt, outer sep=0pt, scale=  0.80] at ( 62.33, 49.65) {GlottDNN};

\node[text=drawColor,anchor=base,inner sep=0pt, outer sep=0pt, scale=  0.96] at ( 63.80,136.43) {\bfseries Nick};
\end{scope}
\end{tikzpicture}